%% file: main.tex
\documentclass[aps,prx,twocolumn,superscriptaddress,citeautoscript,floatfix,showpacs,longbibliography]{revtex4-1}

\usepackage[english]{babel}
\usepackage[utf8]{inputenc}
\usepackage{amsmath}
\usepackage{graphicx}
\usepackage[dvipsnames]{xcolor}
\usepackage{amssymb}
\usepackage{anysize}
\usepackage{amsmath}
\usepackage{physics}
\usepackage{makeidx}
\usepackage{float}
\usepackage{xcolor,colortbl}
\usepackage{listings}
\usepackage{enumitem}
\usepackage[group-separator={,}]{siunitx}
\usepackage{subfigure}
\usepackage{caption}
\usepackage{lipsum}
\usepackage{qcircuit}
\usepackage{xr}
\usepackage[draft]{pdfcomment}

\usepackage[paperwidth=210mm,paperheight=297mm,centering,hmargin=2cm,vmargin=2.5cm]{geometry}
\usepackage{natbib}

\usepackage[all]{hypcap}
\providecommand{\newoperator}[2]{\newcommand*{#1}{\mathop{\mathrm{#2}}\nolimits}}
\newoperator{\sgn}{sgn}
\newoperator{\arctanh}{arctanh}
\newoperator{\argmax}{argmax}
\newoperator{\diag}{diag}

\def\bra#1{\langle#1|}
\def\ket#1{|#1\rangle}

\begin{document}
\title{Quantum Circuits For Two-Dimensional Isometric Tensor Networks}

\author{Lucas Slattery}
\author{Bryan K. Clark}

\affiliation{Department of Physics and IQUIST and Institute for Condensed Matter Theory, University of Illinois at Urbana-Champaign, IL 61801, USA}

\begin{abstract}
The variational quantum eigensolver (VQE) combines classical and quantum resources in order simulate classically intractable quantum states. Amongst other variables, successful VQE depends on the choice of variational ansatz for a problem Hamiltonian. We give a detailed description of a quantum circuit version of the 2D isometric tensor network (isoTNS) ansatz which we call qisoTNS. We benchmark the performance of qisoTNS on two different 2D spin 1/2 Hamiltonians. We find that the ansatz has several advantages. It is qubit efficient with the number of qubits allowing for access to some exponentially large bond-dimension tensors at polynomial quantum cost. 
In addition, the ansatz is robust to the barren plateau problem due emergent layerwise training. We further explore the effect of noise on the efficacy of the ansatz. Overall, we find that qisoTNS is a suitable variational ansatz for 2D Hamiltonians with local interactions.
\end{abstract}
\maketitle

\section{\label{sec:intro} Introduction}
Despite recent advances in quantum computing hardware, quantum computers remain limited in qubits and state coherence. These noisy intermediate-scale quantum (NISQ) computers will for the foreseeable future be unable to perform error correction \cite{Preskill2018}. Without error correction, only short-depth circuits are possible. Amongst the most promising short-depth circuit quantum computing applications are variational quantum algorithms (VQA) \cite{Farhi2014,Peruzzo2014}. VQAs are hybrid classical-quantum algorithms. They leverage quantum computers to simulate and sample from classically intractable states while using classical computers to minimize qubit and circuit depth requirements.

In VQAs, one optimizes a parameterized quantum circuit ansatz with respect to a cost function. Amongst other factors, the success of the optimization relies on the ansatz used. The variational quantum eigensolver (VQE) attempts to find the ground (or excited states) of a Hamiltonian. In VQE, the ansatz can take many different forms each motivated by the problem Hamiltonian and the limitations of the quantum computer. The hardware efficient ansatz used in superconducting qubit arrays is comprised of layers of single qubit gates and native entangling unitaries \cite{Kandala2017}. The unitary coupled cluster ansatz (UCC) is commonly used for simulating molecular Hamiltonians. The UCC maps a Hatree-Fock approximation of the molecular orbitals to qubits. Then the ansatz is formed from a series of tunable interactions between different orbitals forming an efficient ansatz for molecular Hamiltonians \cite{Lee2019}. Other ansatze have also been proposed for use in a variety of systems \cite{Kim,Foss-Feig2020,Khamoshi2021,Liu2019,Grimsley2019,Hadfield2019,Otten}. 

In numerical physics, variational tensor network states have been successful in representing the low-energy states of many-body Hamiltonians. The matrix product state (MPS) and the projected entangled pair state (PEPS) have been used to successfully study gapped 1D and 2D many-body Hamiltonians respectively \cite{Ostlund1995,Verstraete2004}. Recently a restricted form of PEPs, the isometric 2D tensor network state (isoTNS), was introduced \cite{Haghshenas2019,Soejima2019,Zaletel2020,Tepaske2021}. In PEPS, rank 5 tensors are arranged in a grid pattern with the tensor at grid site $(m,n)$ written as $A^{ijklp}_{x_(m,n)}$. Each tensor has a physical degree of freedom $p$ and shares the four index bonds $(i,j,k,l)$  (with size or bond-dimension $\chi$) with the tensor's four neighbors.  See Figure \ref{fig:isoTNS-a} for a graphical depiction of the site tensor. We can write the PEPs representation of a wavefunction as:
\begin{equation}
\label{peps}
     \ket{\Psi}=\sum_{\{x\}}{\mathcal{C}[A^{p_{(0,0)}}_{x_{(0,0)}}...A^{p_{(m,n)}}_{x_{(m,n)}}}]\ket{p_{(0,0)}...p_{(m,n)}}
 \end{equation}
where we have suppressed the bond indices and the contraction is taken over the grid.

IsoTNS gives a canonical form to the projected entangled pair state by adding an additional isometric requirement. When contracting the isoTNS with itself, each of the tensors forms an isometry when contracted towards a central site called an orthogonality center. In this paper, we introduce and benchmark a quantum circuit version of the isoTNS which we call `qisoTNS'. In Section \ref{sec:isotns}, we provide a brief introduction to the classical isoTNS before demonstrating how to construct the quantum circuit version, qisoTNS. In Section \ref{sec:benchmarking-a} and \ref{sec:benchmarking-b}, we benchmark on classical emulators, the performance of qisoTNS for a 2D transverse field Ising (TFI) model and a $J_1-J_2$ Heisenberg model respectively. In Section \ref{sec:benchmarking-c}, we demonstrate how qisoTNS is robust to the barren plateau problem \cite{McClean2018} and in Section \ref{sec:benchmarking-d} we examine the performance of qisoTNS in the presence of simulated noise.

\section{\label{sec:isotns} Isometric 2D Tensor Network Ansatz}
\subsection{\label{sec:iso_intro}Classical isoTNS}
In this section, we first describe the isoTNS before introducing the quantum circuit version, qisoTNS. IsoTNS is a PEPS with the condition that each of the tensors form an isometry when contracted towards a central site called an orthogonality center. Tensors with the same row or column index as the orthogonality center are said to be on the orthogonality hypersurface. For orthogonality center at site $(u,v)$, we can write the isometry condition of tensors not on the hypersurface as:

\begin{equation}
\begin{aligned}
& \sum_{i,j,p}A_{x_{(m,n)}}^{ijklp}A_{x_{(m,n)}}^{\dag ijklp}=\mathbf{1}, \quad \ m < u \ \& \ n < v \\
& \sum_{k,j,p}A_{x_{(m,n)}}^{ijklp}A_{x_{(m,n)}}^{\dag ijklp}=\mathbf{1}, \quad \ m > u \ \& \ n < v \\
& \sum_{i,l,p}A_{x_{(m,n)}}^{ijklp}A_{x_{(m,n)}}^{\dag ijklp}=\mathbf{1}, \quad \ m < u \ \& \ n > v \\
& \sum_{k,l,p}A_{x_{(m,n)}}^{ijklp}A_{x_{(m,n)}}^{\dag ijklp}=\mathbf{1}, \quad \ m > u \ \& \ n > v \\
\end{aligned}
\label{eq:iso_cond}
\end{equation}

where $i,j,k$ and $l$ are the left,top,right and bottom indices respectively and the $p$ index is the physical degree of freedom. Tensors on the orthogonality hypersurface also have an isometry condition when contracted with themselves over all all but one index.

\begin{figure}[!htp]
\subfigure[][]{
\includegraphics[height=52pt]{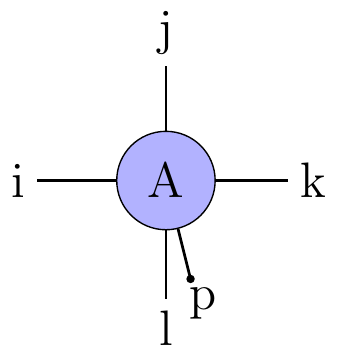}
\label{fig:isoTNS-a}
}
\subfigure[][]{
\includegraphics[height=52pt]{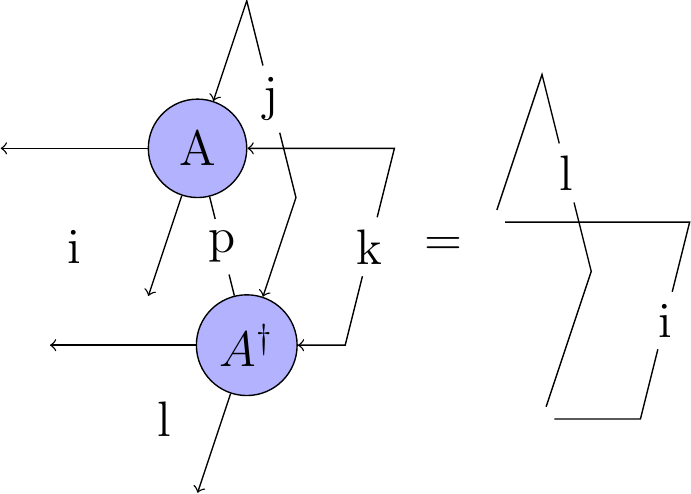}
\label{fig:isoTNS-b}
}
\subfigure[][]{
\includegraphics[height=52pt]{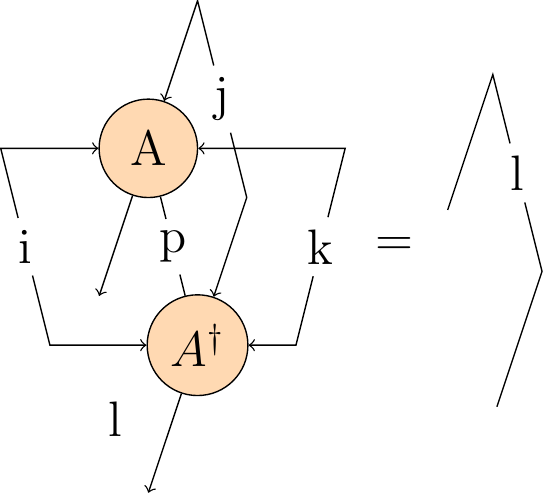}
\label{fig:isoTNS-c}
}
\subfigure[][]{
\includegraphics[width=.71\columnwidth]{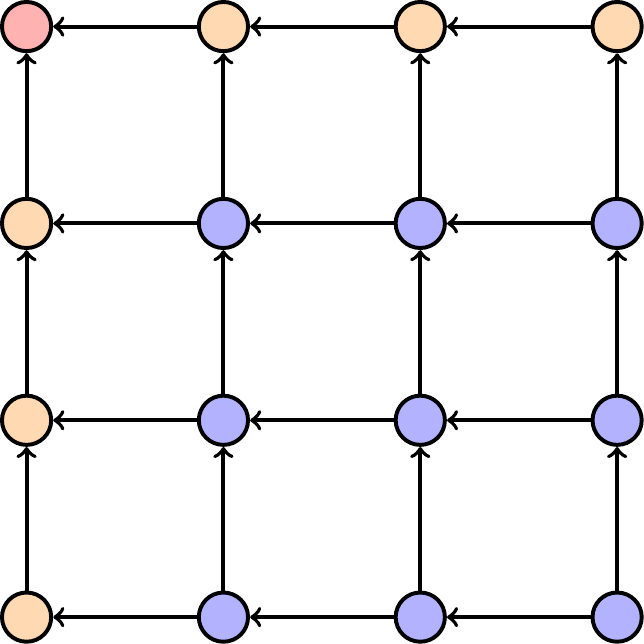}
\label{fig:isoTNS-d}
}
\caption{Tensor diagrams for isoTNS. \subref{fig:isoTNS-a} A rank 5 tensor. The i,j,k and l indices correspond to the left,upper,right and lower bond indices while the p index corresponds to the physical degree of freedom on the site. \subref{fig:isoTNS-b}\subref{fig:isoTNS-c} A tensor diagram of an isometric condition for a tensor in the bulk \subref{fig:isoTNS-b} and on the orthogonality hypersurface \subref{fig:isoTNS-c}. The tensors when contracted with themselves along the necessary indices reduce to the identity. \subref{fig:isoTNS-d} Diagram of an isoTNS. The arrows point in the direction of the isometry. The red tensor is the orthogonality center towards which all isometry conditions point. The orange tensors make up of the orthogonality hypersurface.}
\label{fig:isoTNS}
\end{figure}

Graphically, we can represent an isometric tensor network as a PEPS decorated with a series of arrows all pointing toward an orthogonality center (See Figure \ref{fig:isoTNS-d}). Each tensor within an isoTNS when contracted with itself in the direction of the arrows as in Figure \ref{fig:isoTNS-b} must form the identity. In Figure \ref{fig:isoTNS-c}, we provide a graphical representation for the additional bond index contraction requirement for tensors on the orthogonality hypersurface. Note that the arrows on the isoTNS dictate which bond indices must be contracted.

\subsection{\label{sec:qiso_intro}Quantum IsoTNS}
The qisoTNS is an isoTNS with the orthogonality center fixed
on a corner tensor on the grid which we set to be at site $(u,v)=(0,0)$ on our grid. In this case, all the tensors need to follow the isometry condition
\begin{equation}
 \sum_{k,l,p}A_{x_{(m,n)}}^{ijklp}A_{x_{(m,n)}}^{\dag ijklp}=\mathbf{1}
\label{eq:iso_cond2}
\end{equation}
Unlike in a classical isoTNS, the orthogonality center in a qisoTNS is fixed throughout the entire algorithm.

The qisoTNS ansatz is a holographic quantum circuit \cite{Kim2017}; although not explicitly described, ref.~\cite{Foss-Feig2020} first mentions the possibility that isoTNS could be implemented as holographic circuits. In a holographic quantum circuit, a 2D system is simulated with a 1D array of qubits evolving in time - i.e. the second dimension is the circuit evolution. For a holographic quantum circuit, this means the number of qubits needed to simulate a 2D system scales with the width and not the area of the system. In qisoTNS, similarly to the holographic MPS quantum circuit (and its quasi-2D variant), site indices are measured not just on different qubits but on the same qubit at different times during the circuit operation \cite{Foss-Feig2020,Liu2019}. 

In order to construct the qisoTNS, we use unitary matrices $U$ to represent the tensors of the network.  To do this, we split the indices of the matrix to form a rank 6 tensor, $U^{(ijt)(pkl)}$. In this notation,  $ijt$ ($pkl$) corresponds to the rows (columns) of the matrix.  By fixing the $t$ index to be zero, the unitary matrix now represents a rank 5 tensor that (due to unitarity) satisfies the isometry condition of Eq. \ref{eq:iso_cond2} where U now corresponds to $A^{ijklp}$. 

On a circuit, the $i,j,t$ indices correspond to input wires for the unitary block and the $k,l,p$ indices correspond to the output wires. 
Multiple qubits/wires can be used to represent each index of $U$.  The $t$ wires always take $|0\rangle$ as input fixing that index and the $p$ wires are always measured. 

\begin{figure}[!htp]
\subfigure[][]{
\includegraphics[height=52pt]{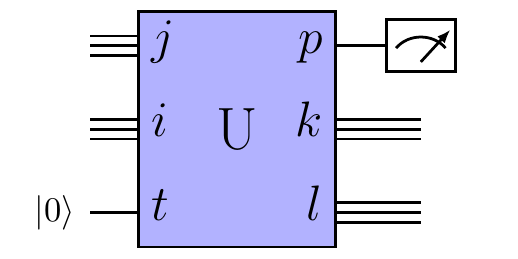}
\label{fig:circ_comp-a}
}
\subfigure[][]{
\includegraphics[height=52pt]{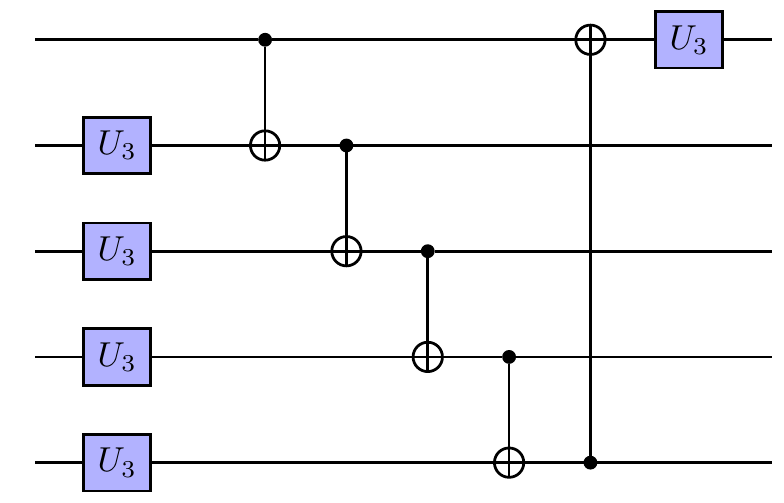}
\label{fig:circ_comp-b}
}
\subfigure[][]{
\includegraphics[width=.88\columnwidth]{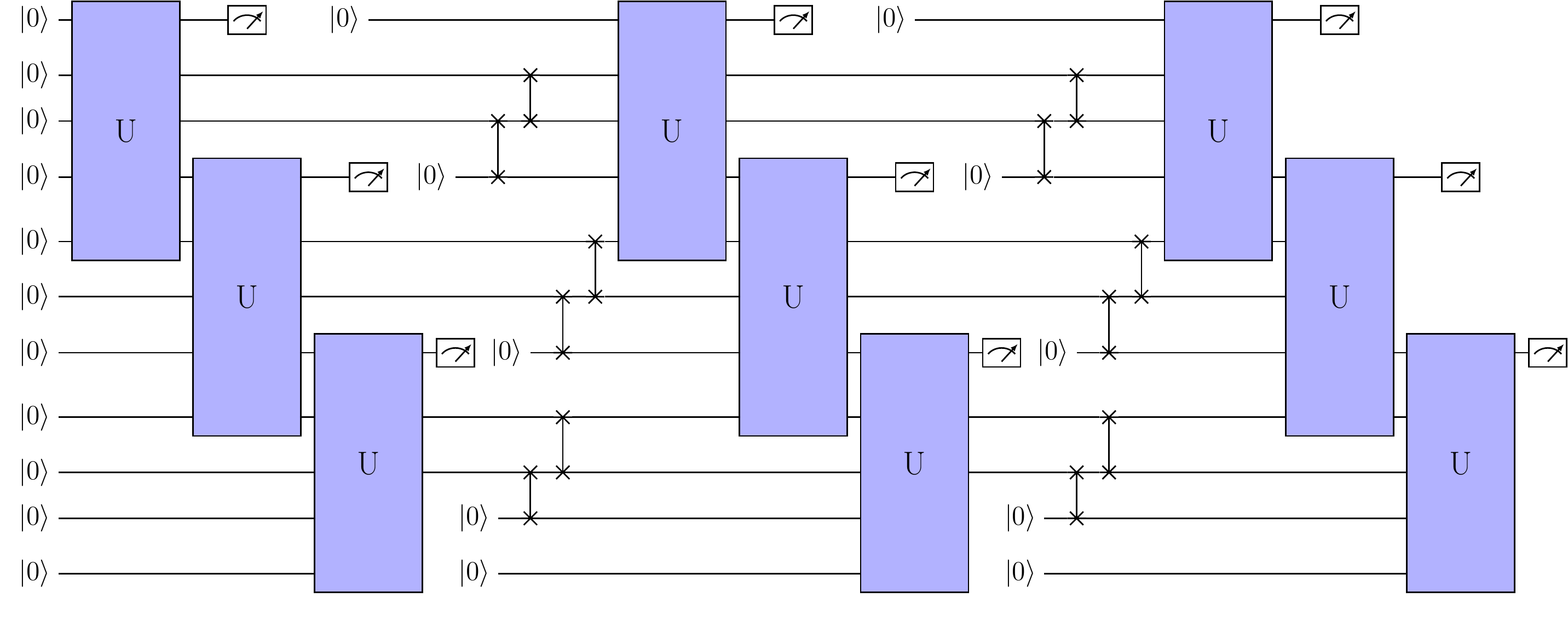}
\label{fig:circ_comp-c}
}
\caption{Circuit diagrams for qisoTNS. \subref{fig:circ_comp-a} Circuit diagram for unitary block $U$ representing a site tensor. Each unitary matrix can be mapped to a rank 6 tensor. \subref{fig:circ_comp-b} For our benchmarking tests, each unitary block $U$ representing a site tensor is made up of $n_{bl}$ layers of the pattern of $U_3$ and CNOT gates shown here. \subref{fig:circ_comp-c} A circuit diagram for a $3\times 3$ spin $1/2$ qisoTNS with $n_{bq}$=2. The t index in $A^{ijklpt}$ is contracted with an incoming qubit to create the rank 5 tensor. Swap networks are used to connect the outgoing  horizontal bonds with the appropriate incoming bonds of the tensor in the column over.}
\label{fig:circ_comp}
\end{figure}

These unitary matrices represented as circuit blocks are used to create the qisoTNS. The qisoTNS ansatz is constructed by layering unitary blocks in a grid pattern as in Figure \ref{fig:circ_comp-c}.  They are arrayed so that the $l$ (down) index of a unitary block connects with the $j$ (up) index of the unitary block in the row below it while the $k$ (right) index connects via swap gates with the $i$ (left) index of the unitary matrix in the column to the right. Measuring the qubit corresponding to the p (physical) index immediately after a unitary matrix is equivalent to sampling from the index in the measurement basis allowing one to compute operator expectation values. 
Note that each of the qisoTNS rows and columns when viewed separately can be viewed as an MPS in canonical form where the canonical site is on the left or top edge for our choice of orthogonality center.  QisoTNS can be seen as a several MPS circuits stacked vertically or horizontally.

There are several important ansatz hyper-parameters that determine the expressiblity of a given qisoTNS circuit. Firstly, the number of qubits $n_{bq}$ used per bond determines the bond dimension of the ansatz, $\chi = 2^{n_{bq}}$.  Generically one could have separate horizontal and vertical bond-dimensions but throughout this paper we make the simplifying assumption that they are the same. As an example, the unitary matrix in Figure $\ref{fig:circ_comp-a}$ $n_{bq}=3$ while the unitaries in Figure $\ref{fig:circ_comp-c}$ have $n_{bq}=2$.   Secondly, $U$ can be an arbitrary circuit block (as any such circuit satisfies the isometry conditions of Equation \ref{eq:iso_cond2}). To represent a generic unitaries, $U$ (and hence a completely arbitrary tensor) generically requires $O(2^{2n_{bq}+n_{p}})$ gate depth where $n_p$ is the number of qubits representing the physical degree of freedom and therefore we don't get an exponential speed up for generic tensors. In this paper, we use a hardware efficient ansatz generated from $n_{bl}$ layers of Figure \ref{fig:circ_comp-b}. Using this $U$, the number of qubits required and the circuit depth both scale as $O(n_{bq})$ or $O(\log{\chi})$. This gives access to a subset but not all of the exponentially large bond dimension with linear quantum resources.

When compared with classical isoTNS, qisoTNS has several differences. In isoTNS, in order to efficiently measure an operator expectation value, the operator must be inside the orthogonality hypersurface. When the operator is on the hypersurface, contracting isoTNS reduces the problem of calculating expectation values on an MPS. The computational cost of calculating expectation values therefore is dominated by the cost moving the orthogonality center. Currently, for a spin 1/2 system, the cost of moving the orthogonality scales as $O(\chi^{7})$ when using the so called 'Moses Move' as in Ref. \cite{Zaletel2020} or $O(\chi^{8})$ when using the QR decomposition direct minimization method as in Ref. \cite{Haghshenas2019}.  
In qisoTNS, the contraction of the tensor network is achieved by the circuit and the observables computed don't need to be on the orthogonality center. The orthogonality center does not (and cannot) move. Instead, the cost of calculating expectation values is dominated by the cost of running the qisoTNS circuit.

In isoTNS, because we have to move the orthogonality center, we are computing expectation values on a series of tensor networks which are approximately equal introducing an additional error beyond just the variational approximation.  There is no such additional approximation in qisoTNS because the orthogonality center remains fixed.  In addition, by measuring the qisoTNS one samples directly from the square of the wave-function whereas we are unaware of any way to sample this probability distribution in classical isoTNS efficiently. Interestingly, in fact the nature of the causality cone of this circuit allows us to generate directed conditional probabilities where the outcome of measurements on tensors further from the orthogonality center in qisoTNS is dependent on the closer tensors but not vice versa. While in this paper we study the isometric version of PEPS, in general one can construct much more general isometric tensor networks using holographic quantum circuits. See Appendix \ref{sec:AppA} for an example. As far as we know, computing expectation values using these more complicated networks cannot be done efficiently classically.

\begin{figure*}[!htp]
\centering
\subfigure[][]{
\includegraphics[width=152pt]{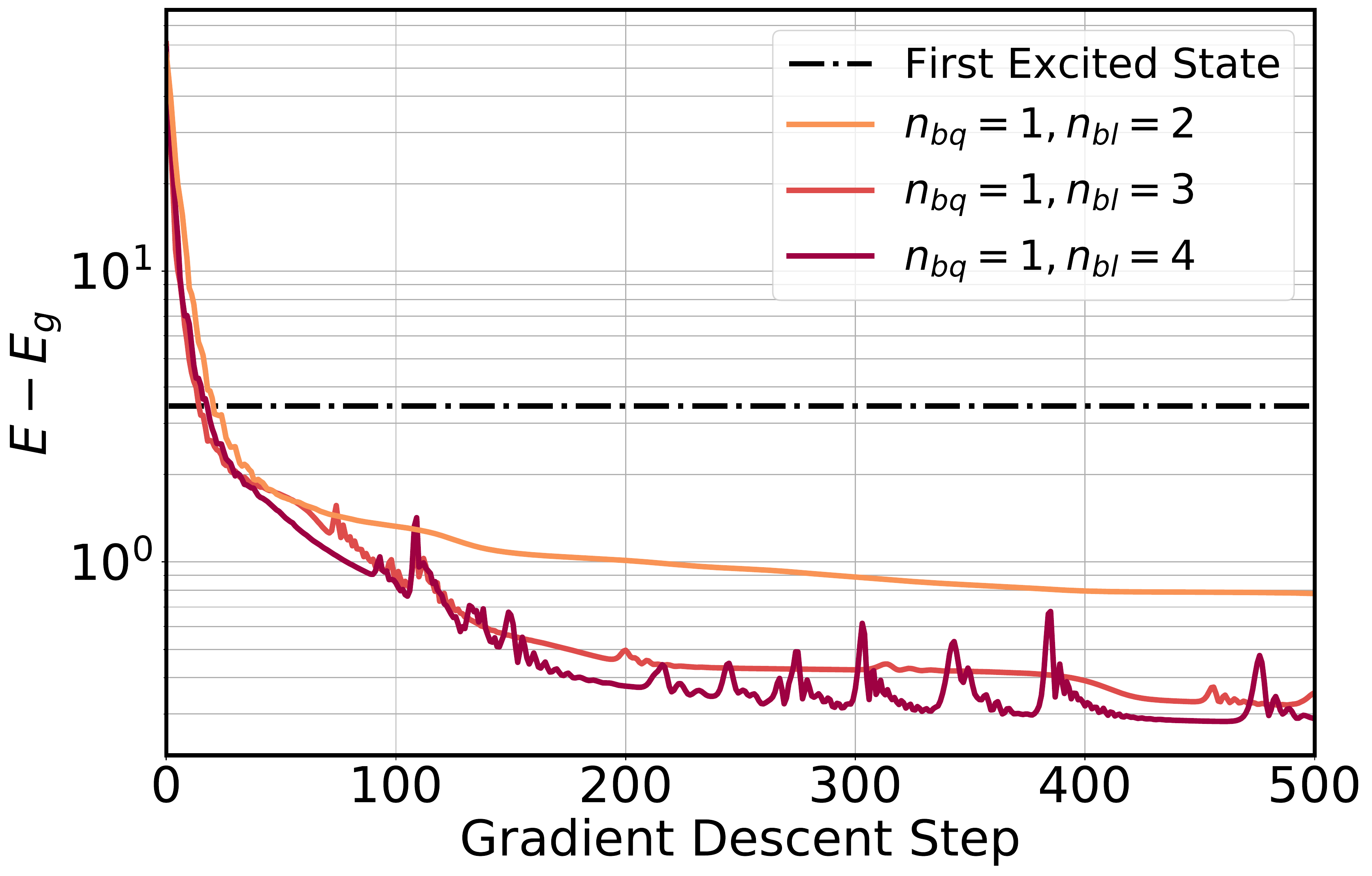}
\label{fig:TFIexample-a}
}
\subfigure[][]{
\includegraphics[width=152pt]{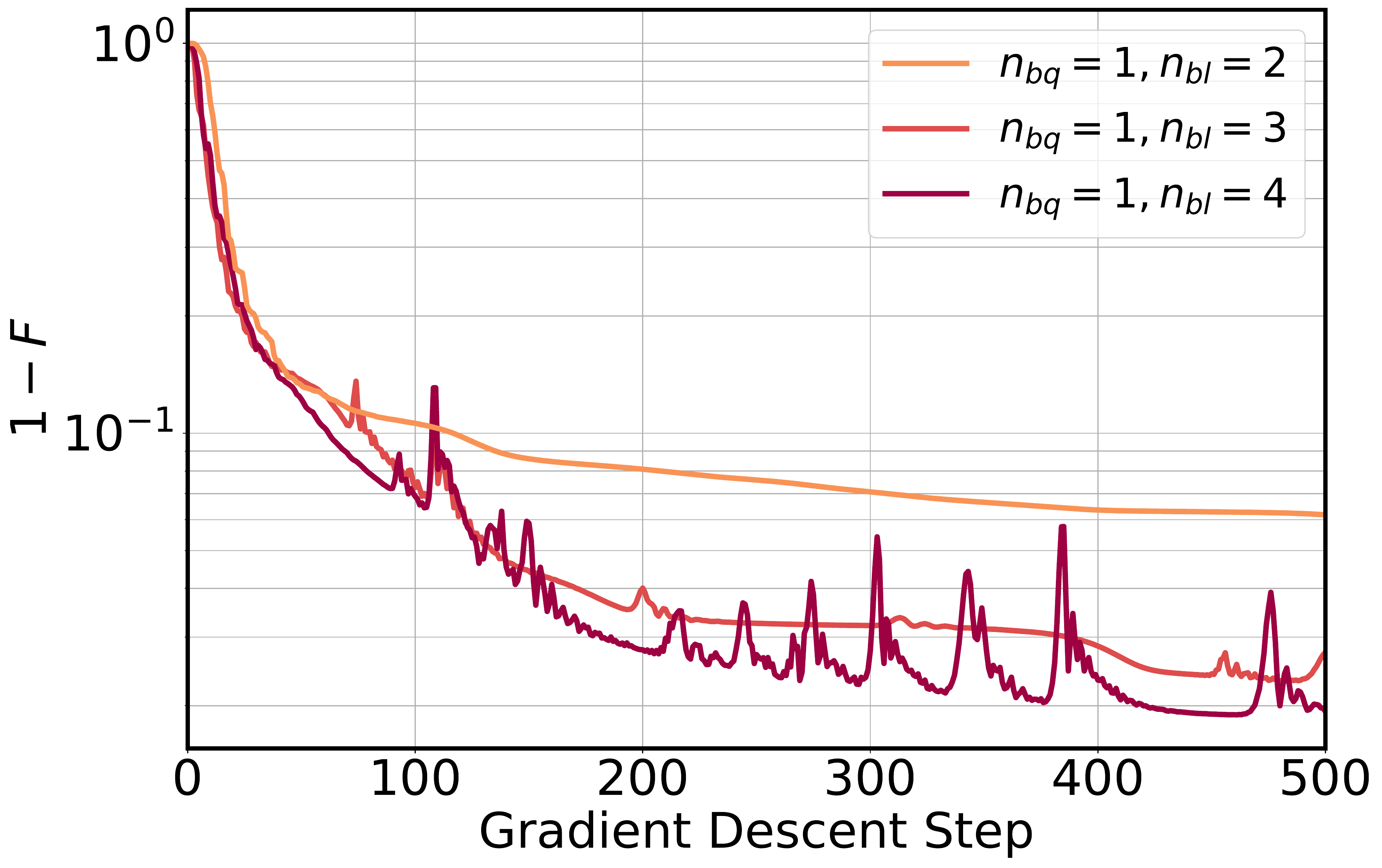}
\label{fig:TFIexample-b}
}
\subfigure[][]{
\includegraphics[width=152pt]{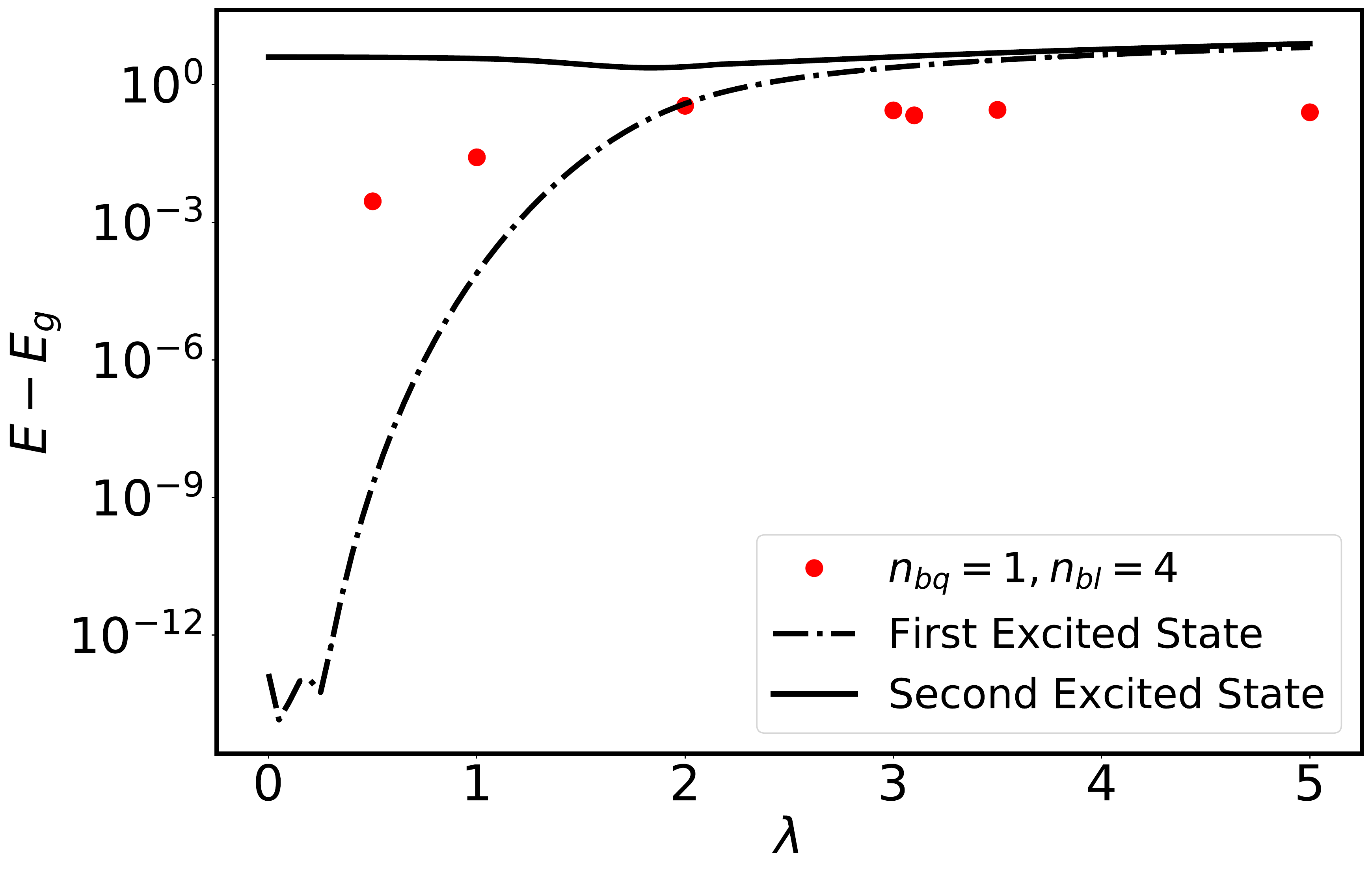}
\label{fig:TFIexample-c}
}
\caption{VQE plots for the 2D TFI model. \subref{fig:TFIexample-a}\subref{fig:TFIexample-b} VQE optimization using classically computed exact expectation values for a 4x4 2D TFI model with $\Delta=1$ and $\lambda=3.5$ for ground state energy difference from the ground state \subref{fig:TFIexample-a} and ground state fidelity error \subref{fig:TFIexample-b}. We use ansatze with $n_{bq}=1$ and $n_{bl}$ = 2,3 and 4 layers of $U_3$ and CNOT gates as in Figure \ref{fig:circ_comp-b}. \subref{fig:TFIexample-c} Energy reached after 500 AMSgrad steps for a $n_{bq}=1$ and $n_{bl}=4$ ansatze.  Note that the first excited state is degenerate in the thermodynamic limit  for $\lambda \lesssim 3.05$.}
\label{fig:TFIexample}
\end{figure*}

\begin{figure*}[!htp]
\centering
\subfigure[][]{
\includegraphics[width=230pt]{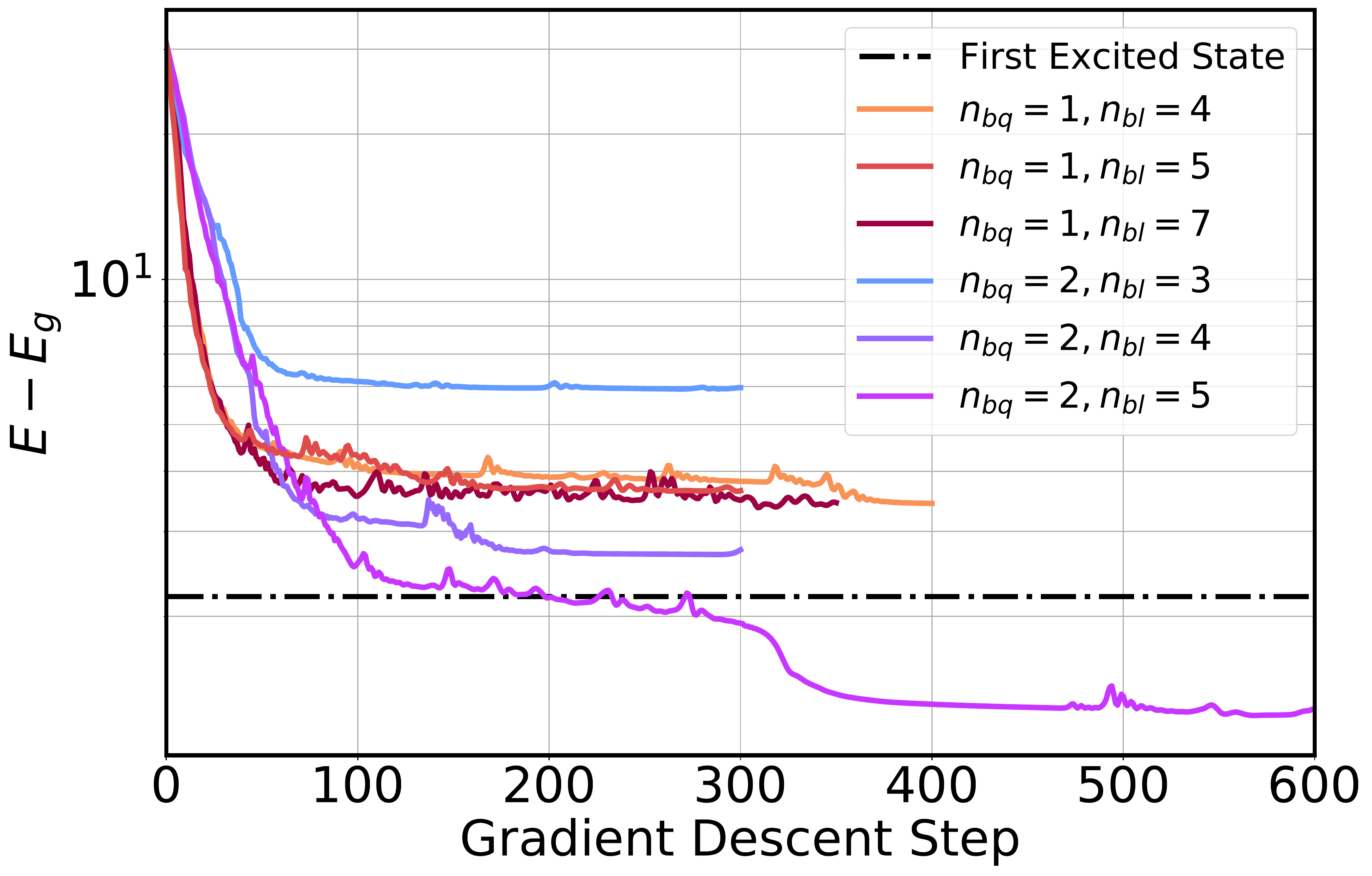}
\label{fig:j1j2example-a}
}
\subfigure[][]{
\includegraphics[width=230pt]{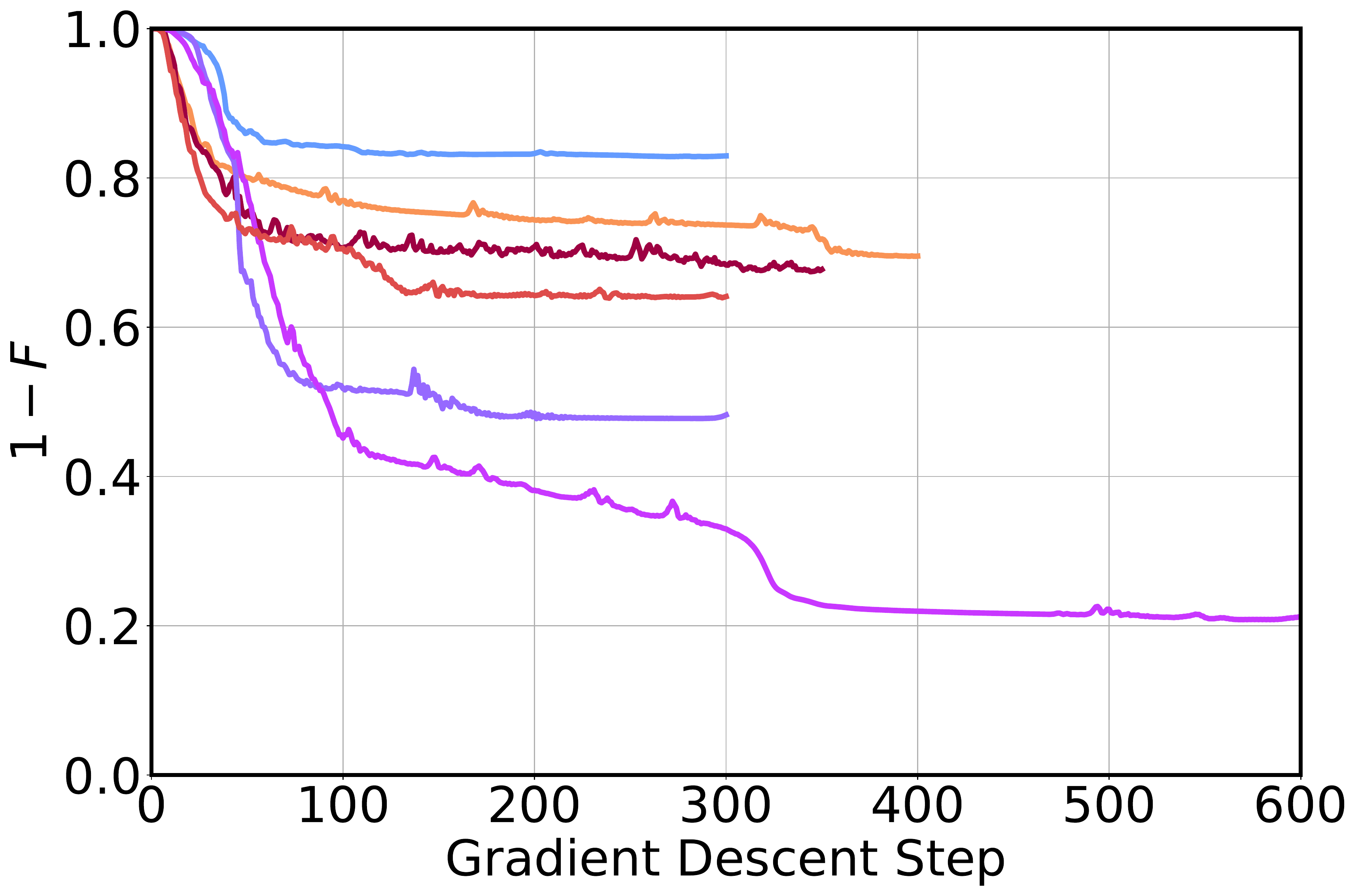}
\label{fig:j1j2example-b}
}
\caption{VQE optimization of a 4x4 $J_1$-$J_2$ Heisenberg model with $J_1=1$ and $J_2=0.5$. \subref{fig:j1j2example-a} Energy above ground state and \subref{fig:j1j2example-b} fidelity error vs. AMSgrad gradient steps for qisoTNS with $n_{bq}=1,2$ and varying $n_{bl}$.}
\label{fig:j1j2example}
\end{figure*}

We can also contrast qisoTNS with other quantum tensor networks. The deep multi-scale entanglement renormalization ansatz (DMERA) implements the MERA tensor network as a holographic quantum circuit \cite{Kim}. Ref.~\cite{Foss-Feig2020,Liu2019,Huggins2019} describes holographic matrix-product states with a recently posted preprint \cite{Haghshenas} arguing that these holographic quantum MPS produce asymptotically more efficient parameterization for quantum ground states then classical MPS. MPS are well suited for one-dimensional systems; qisoTNS differs primarily in its ability to access two-dimensional systems.  In ref.~\cite{Liu2019}, they introduce a variational ansatz, qPEPS, which works in two-dimensions by treating the entire second dimension as a single super-site. This will require bond-dimensions in the width which grows exponentially with system size. An alternative construction of PEPS in ref.~\cite{Schwarz2012} produces an injective PEPS on a quantum computer in time proportional to the inverse of the spectral gap of its parent Hamiltonian. As contracting a general PEPS is computationally intractable even for a quantum computer, the subset of PEPS considered in these works must generically be costly \cite{Schuch2007}. We work instead with a subset of PEPS, the isoTNS which can be efficiently contracted on a quantum computer. Algorithms to generate tensor networks from photon scattering have also been considered \cite{Schon2005,Schon,Lindner,Schwartz2021,Pichler2017,Xu2018}

\section{\label{sec:benchmarking} Benchmarking qisoTNS with VQE}

In this section we benchmark the performance of the ansatz on two different 2D spin 1/2 Hamiltonians on a square lattice using VQE. We look at ansatze with $n_{bq}=1$ and $2$. Each unitary block representing a site tensor is made up of $n_{bl}$ layers of parameterized $U_3$ gates and CNOT gates as in Figure \ref{fig:circ_comp-b}. For both the TFI and J1-J2 model, we optimize our ansatz using the AMSgrad optimizer and exact tensor network simulations of the quantum circuit starting from randomly initialized parameters \cite{Reddi,Kingma}. For the J1-J2 model, we look at the noise resilience of the energy and of $\sigma^{z}_{i}\sigma^{z}_{j}$ correlators at different circuit depths using noisy classically simulated Qiskit circuits \cite{Aleksandrowicz2019}. In addition, we look at the optimization of a $4 \times 4$ TFI model under exact and noisy conditions.
\subsection{\label{sec:benchmarking-a} 2D TFI Model}
The 2D TFI model on a square lattice is defined by:
\begin{equation}
\hat{H}=\lambda \sum_{i}\sigma^{x}_{i}+\Delta\sum_{\langle i,j \rangle}\sigma^{z}_{i}\sigma^{z}_{j}
\end{equation}

In the infinite size limit, the ground state undergoes a second order phase transition from ferromagnetic ($\Delta<0$) or antiferromagnetic ($\Delta>0$) phase  to the paramagnetic phase which occurs at $\frac{\lambda}{|\Delta|}\simeq3.044$ \cite{Rieger1999}. The ground state of the (anti)ferromagnetic phase is twofold degenerate due to the $\mathbf{Z}_{2}$ symmetry of the Hamiltonian. At the critical point $\lambda_{c}$, a gap opens up and gapped phases generically need lower $\chi$ to represent them. 

We optimize the qisoTNS ansatz on a 4x4 TFI system for $\Delta=1$ and variable $\lambda$. In Figure \ref{fig:TFIexample-a}\subref{fig:TFIexample-b}, we show a VQE optimization of qisoTNS with $\lambda$=3.5 in the gapped paramagnetic phase. We fix $n_{bq}=1$ ($\chi=2$) while letting $n_{bl}$=2,3 and 4 for our ansatz. The combined number of $U_3$ and CNOT gates in the ansatz ranges from 192 for $n_{bl}=2$ to 384 for $n_{bl}=4$. After just 500 AMSgrad gradient steps, all of the optimizations reach well below the first excited state with fidelities ($F=|\bra{\Psi_{qiso}}\ket{\Psi_{ground}}|^2$) greater than 0.93 with the $n_{bl}=4$ ansatz reaching a final $F$ of 0.981. In Figure \ref{fig:TFIexample-c}, we plot as a function of $\lambda$, the minimum energy reached after 500 AMSgrad gradient steps for a $n_{bq}=1, n_{bl}=4$ ansatz. For $\lambda  > \lambda_c$, the ansatz reaches well below the first excited state. For $\lambda < \lambda_c$, the ground state is nearly degenerate with the first excited state (in the infinite size limit they would be degenerate) and the ansatz reaches well below the second excited state.

\subsection{\label{sec:benchmarking-b} 2D \texorpdfstring{$J_1-J_2$} $ Heisenberg Model  }

The 2D $J_1-J_2$ Heisenberg model on a square lattice is defined by:

\begin{equation}
\hat{H}=J_1\sum_{<i,j>}\vec{\mathbf{\sigma_i}}\cdot\vec{\mathbf{\sigma_j}}+J_2\sum_{<<i,j>>}\vec{\mathbf{\sigma_i}}\cdot\vec{\mathbf{\sigma_j}}
\end{equation}

In the infinite size limit with $J_1,J_2>0$, the ground state exhibits three phases. For $\frac{J2}{J1}\lesssim0.4$, the ground state is in the antiferromagnetic (AFM) Neel ordered phase. For $\frac{J2}{J1}\gtrsim0.6$, the ground state is in the collinear ordered phase. In between the the AFM Neel and collinear phases, the ground state is in a quantum paramagnetic (QPM) phase with long range entanglement \cite{Sirker2006,Schmidt2008}. 
We apply qisoTNS to the intermediate QPM phase with $J_1=1$ and $J_2=0.5$ on a 4x4 square lattice. In Figure \ref{fig:j1j2example}, we show a VQE optimization for these model parameters. We examine the ansatze performance for $n_{bq}=1$ and $2$ with varying $n_{bl}$. We find that the $n_{bq}=2,n_{bl}=4,5$ ansatz outperforms all of the $n_{bq}=1$ ansatze tested. We conjecture that the $n_{bq}=2$ ansatz does a better job of capturing the long range entanglement of the QPM phase.

\subsection{\label{sec:benchmarking-c} QisoTNS and the Barren Plateau}

\begin{figure*}[!htp]
\centering
\subfigure[][]{
\includegraphics[width=152pt]{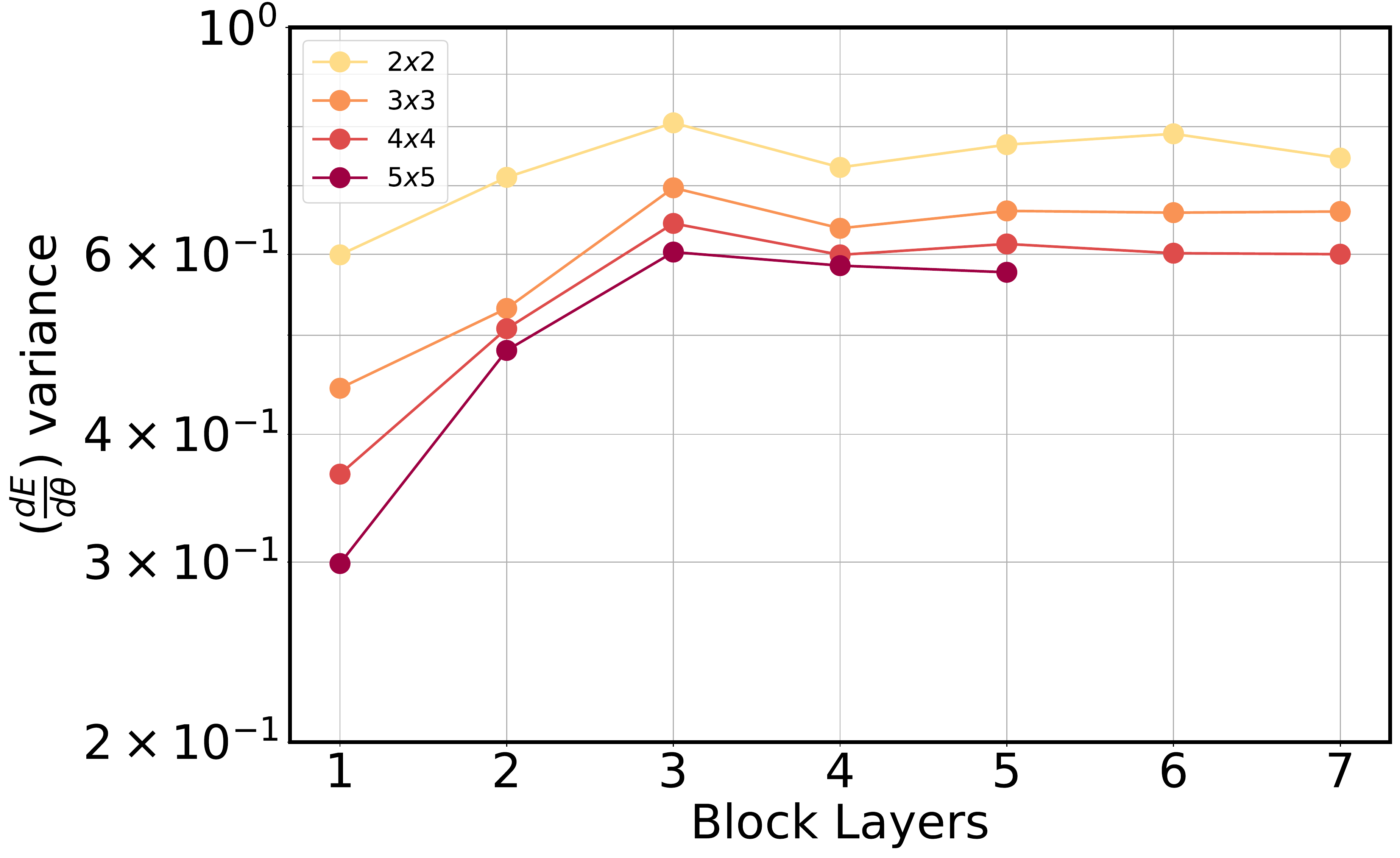}
\label{fig:bp-a}
}
\subfigure[][]{
\includegraphics[width=152pt]{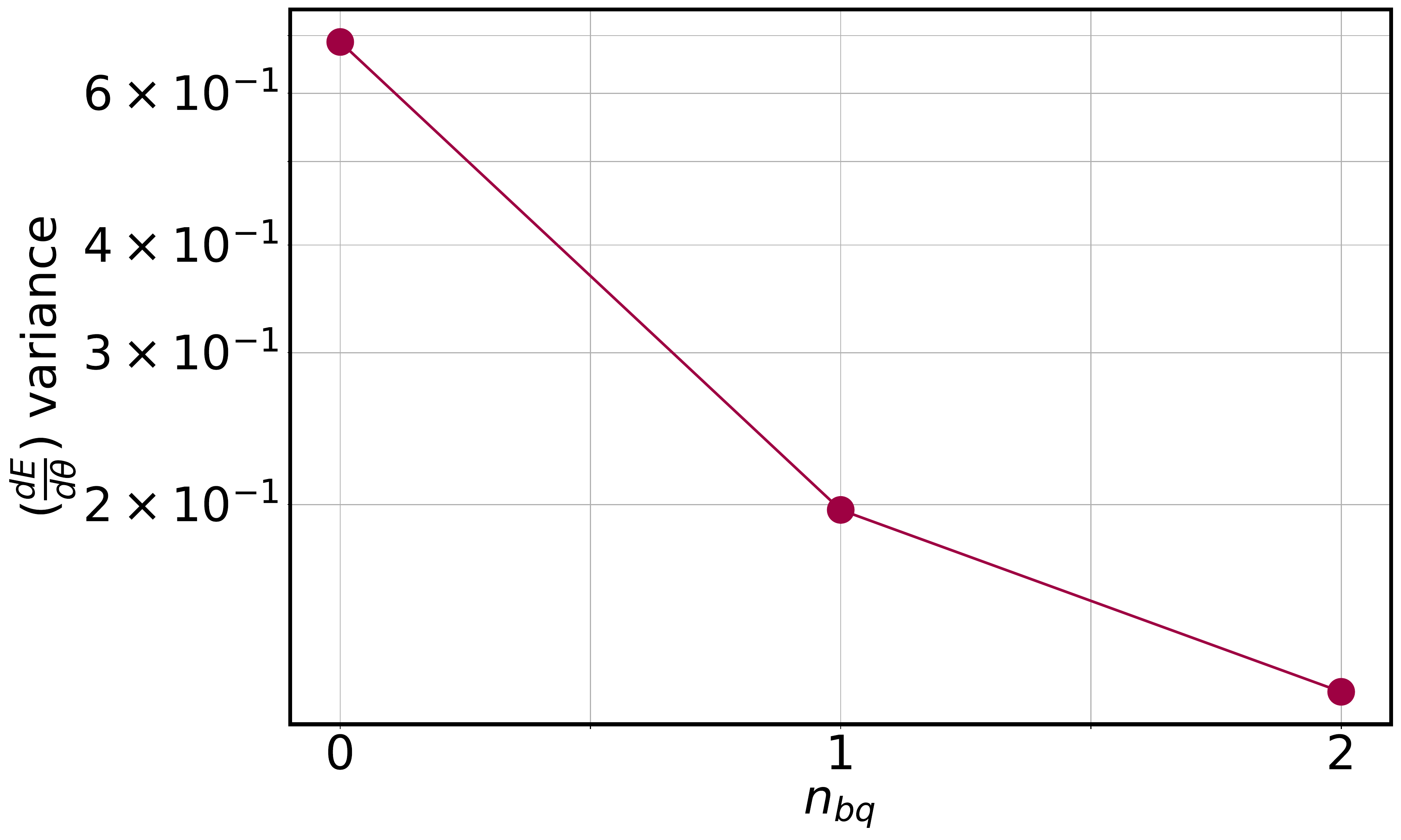}
\label{fig:bp-b}
}
\subfigure[][]{
\includegraphics[width=152pt]{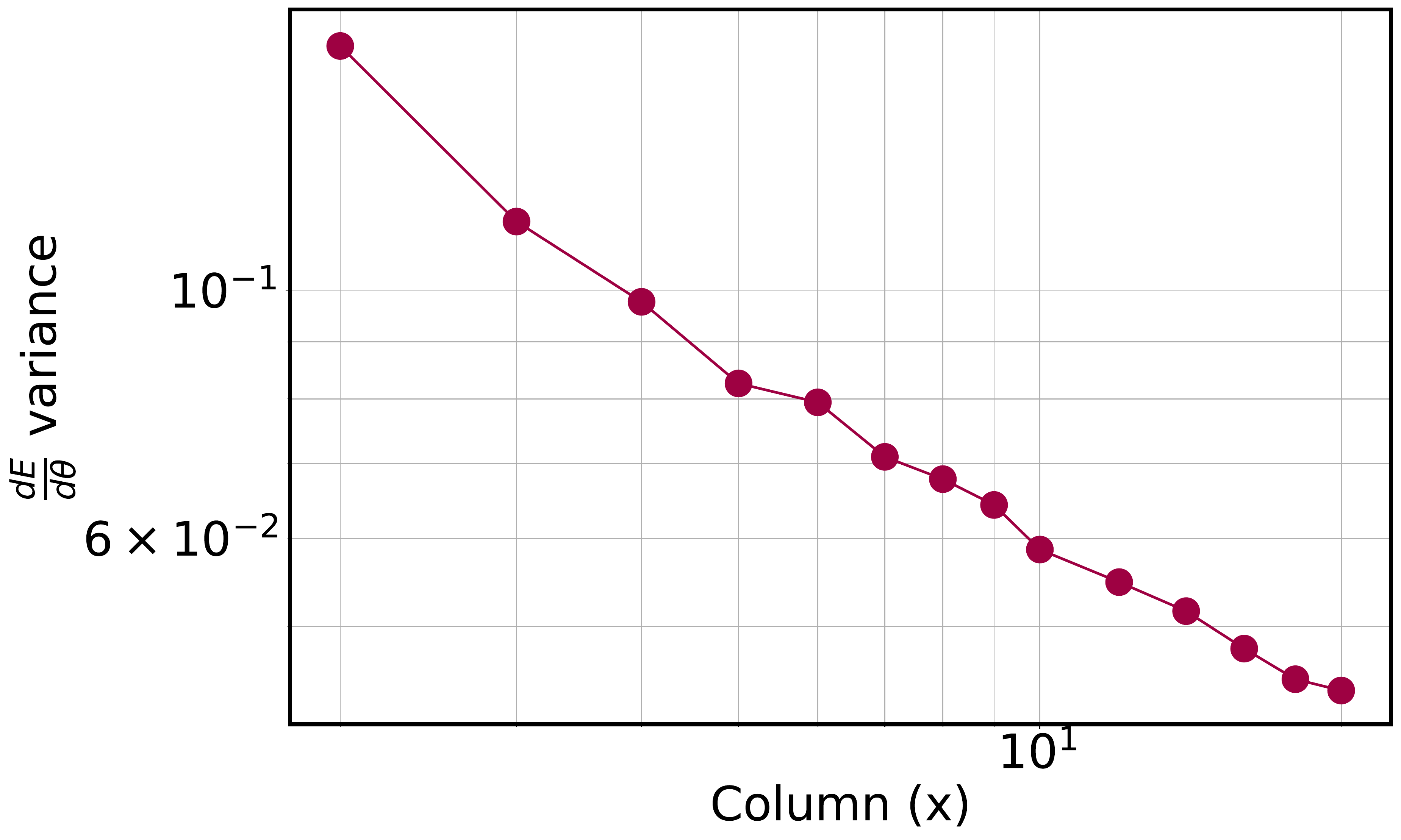}
\label{fig:bp-c}
}
\caption{All plots in this figure use a 2D TFI Hamiltonian with $\Delta=1$ and $\lambda=3.5$ with classically computed exact expectation values for randomly initialized qisoTNS ansatz. \subref{fig:bp-a} Plot of the variance of the gradient vs. increasing $n_{bl}$ in the qisoTNS ansatz for different system size and constant bond qubits, $n_{bq}=1$. Neither increasing the number of layers nor the system size leads to a vanishing variance. \subref{fig:bp-b} Plot of the variance of the gradient vs. number of bond qubits for a $4 \times 4$ qisoTNS with constant block layers, $n_{bl}=4$. \subref{fig:bp-c} Plot of the variance of the gradient of a single $\sigma^z_{3,x}\sigma^z_{3,x+1}$ term acting on last row (row 3) and variable column ($x$) on a $4 \times 20$,$n_{bq}=1$ and $n_{bl}=4$ qisoTNS ansatz.}
\label{fig:bp}
\end{figure*}

For a large class of quantum circuits, the gradient of the objective function with respect to any parameter is expected to decrease exponentially with system size and depth \cite{McClean2018}.This is the so-called barren plateau problem. Additionally, entanglement and noise-induced barren plateaus may exist in quantum circuits \cite{Wang,Marrero}. Multiple methods for alleviating the barren plateau problem including improved ansatz design \cite{Lee2019,Pesah,Zhao,Duncan2020,Zhang2020}, using improved cost functions \cite{Cerezo,Uvarov2021,Patti} and ansatz training and initialization strategies \cite{Slattery,Grant2019,Patti,Volkoff} have been proposed. In this section, we demonstrate that qisoTNS ansatz is robust to the barren plateau.

In qisoTNS, the barren plateau only exists with respect to the number of bond qubits, $n_{bq}$ for local cost functions (See Figure \ref{fig:bp-b}). Fortunately, the bond dimension scales as $\chi=2^{n_{bq}}$ meaning that gradient decays linearly with bond dimension. In Figure \ref{fig:bp-a}\subref{fig:bp-b}, we show plots of the gradient variance with respect to the number of block layers and with respect to the number of bond qubits. Contrasted with the predicted behavior of the barren plateau, we do not see an exponential decay in gradient variance with respect to the number of layers or with increasing system size \cite{McClean2018}. This circuit avoids the typical barren plateaus because Hamiltonian terms acting on sites in the shallow part of the circuit always contribute large variance despite the fact that some of the Hamiltonian terms deep in the circuit do decay quickly (see Figure \ref{fig:bp-c}).

In order to better understand this behavior, consider a $N \times N$ qisoTNS and a Hamiltonian $\hat{H}=\sum{_i}h_i$ where the terms $h_i$ are local operators on the qisoTNS sites. Due to the isometry condition of the qisoTNS, in order to evaluate the gradient or energy of a term $h_i$ on a qisoTNS, one only has to run the circuit up to the furthest site (from the orthogonality center) which $h_i$ acts on. Therefore, for Hamiltonians with operators on all sites, the variance of the gradient for Hamiltonian terms is expected to decay with distance from orthogonality center. This fact ensures that some of the circuit parameters remain trainable regardless of system size allowing us to employ layerwise training strategies to ensure all parameters can be effectively trained. Layerwise training is commonly used in machine learning and has recently been proposed to alleviate the barren plateau problem in quantum circuits \cite{Hettinger,Hinton,Fahlman,Bengio,Skolik2021,Lyu}. In layerwise training, one first trains a shallow quantum circuit before iteratively adding layers and continuing to train the circuit. By pretraining the shallow circuit, one can avoid a random initialization of the deep parts of the quantum circuit leading to exponentially vanishing gradients \cite{Skolik2021}. 

\begin{figure}[!h]
\includegraphics[width=\linewidth]{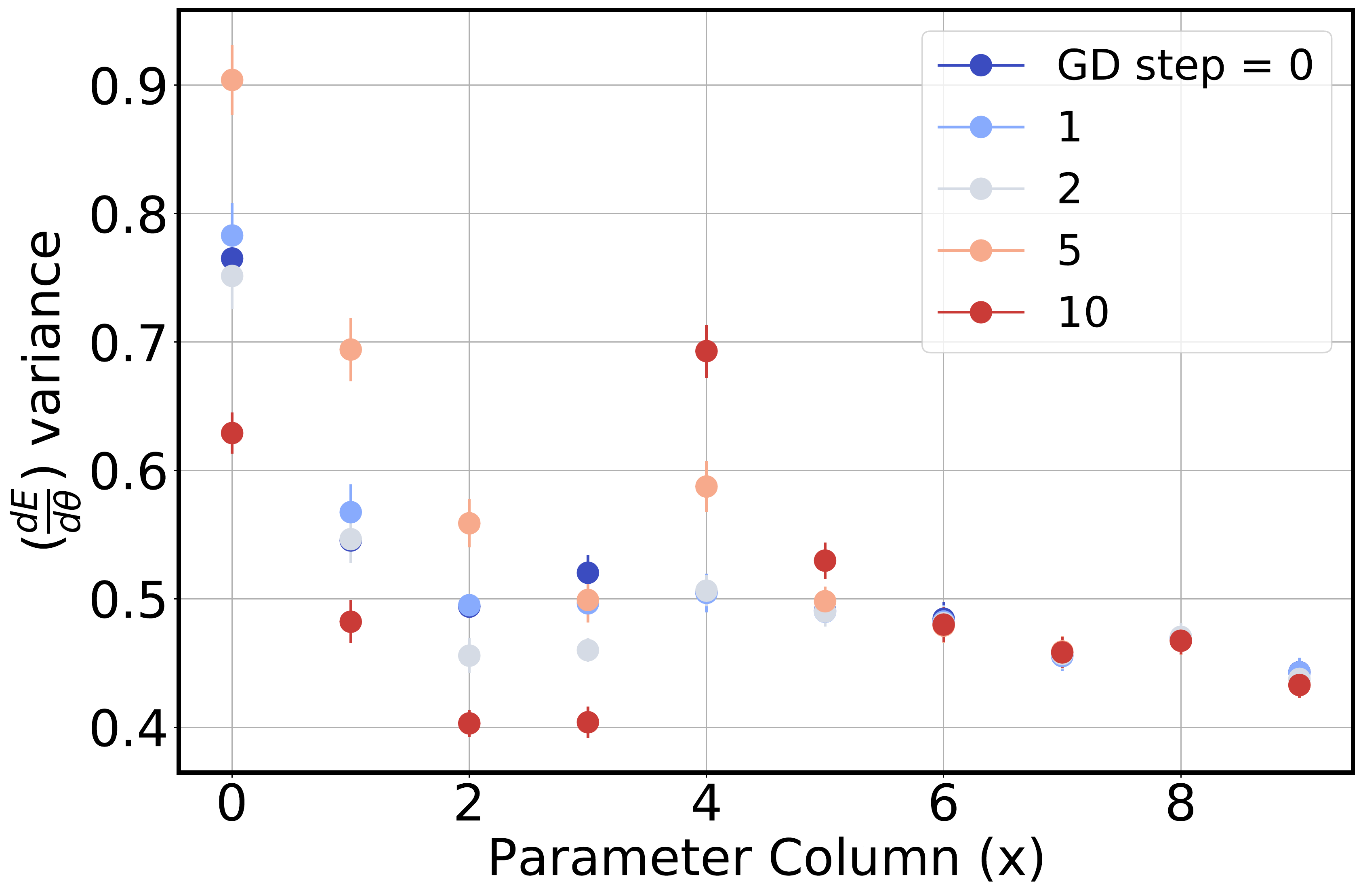}
\caption{Plot of the gradient variance for the full $4\times10$ 2D TFI model with $\Delta=1.0$ and $\lambda=3.5$ as a function of column $x \in [0,9]$. We optimize only on the upper-right $4  \times 4$ tensors with respect to Hamiltonian terms which are completely contained in this $4 \times 4$ block (i.e. on a $4 \times 4$ Hamiltonian). 
We see that the untrained parameters on the border of the trained region (columns 4 and 5) see an increase in gradient variance and hence trainability once the prior layers have been well trained. The data shown is an average of 30 randomly initialized ansatze.} 
\label{fig:pretraining}
\end{figure}

In Figure \ref{fig:pretraining}, we demonstrate how pretraining a smaller qisoTNS can make a larger qisoTNS more trainable.  We demonstrate this on the 2D TFI model with $\Delta=1$ and $\lambda =3.5$. On a $4 \times 10$ qisoTNS, we optimize only the unitary blocks corresponding to the $4 \times 4$ grid closest to the orthogonality center on Hamiltonian terms which are completely contained within it.  This is essentially training  $4 \times 4$ TFI model.   We observe an increase in gradient variance for the parameters immediately bordering the trained region (i.e. column 4 and 5); this indicates that training the shallower sites increases the trainability of these deeper sites alleviating the problem with the barren plateau. 

We note while we have pretrained our $4 \times 10$ ansatz with a $4 \times 4$ Hamiltonian in order to demonstrate the effect of pretraining it is not necessary for qisoTNS. 
Training with the full Hamiltonian is inherently layerwise training in qisoTNS (as well as in other holographic circuits) as the gradients of the parameters shallower in the circuit are unaffected by Hamiltonian terms deeper in the circuit while the reverse is not true.

\subsection{\label{sec:benchmarking-d} Qiskit Noise Tests}

\begin{figure*}[!htp]
\subfigure[][]{
\includegraphics[width=152pt]{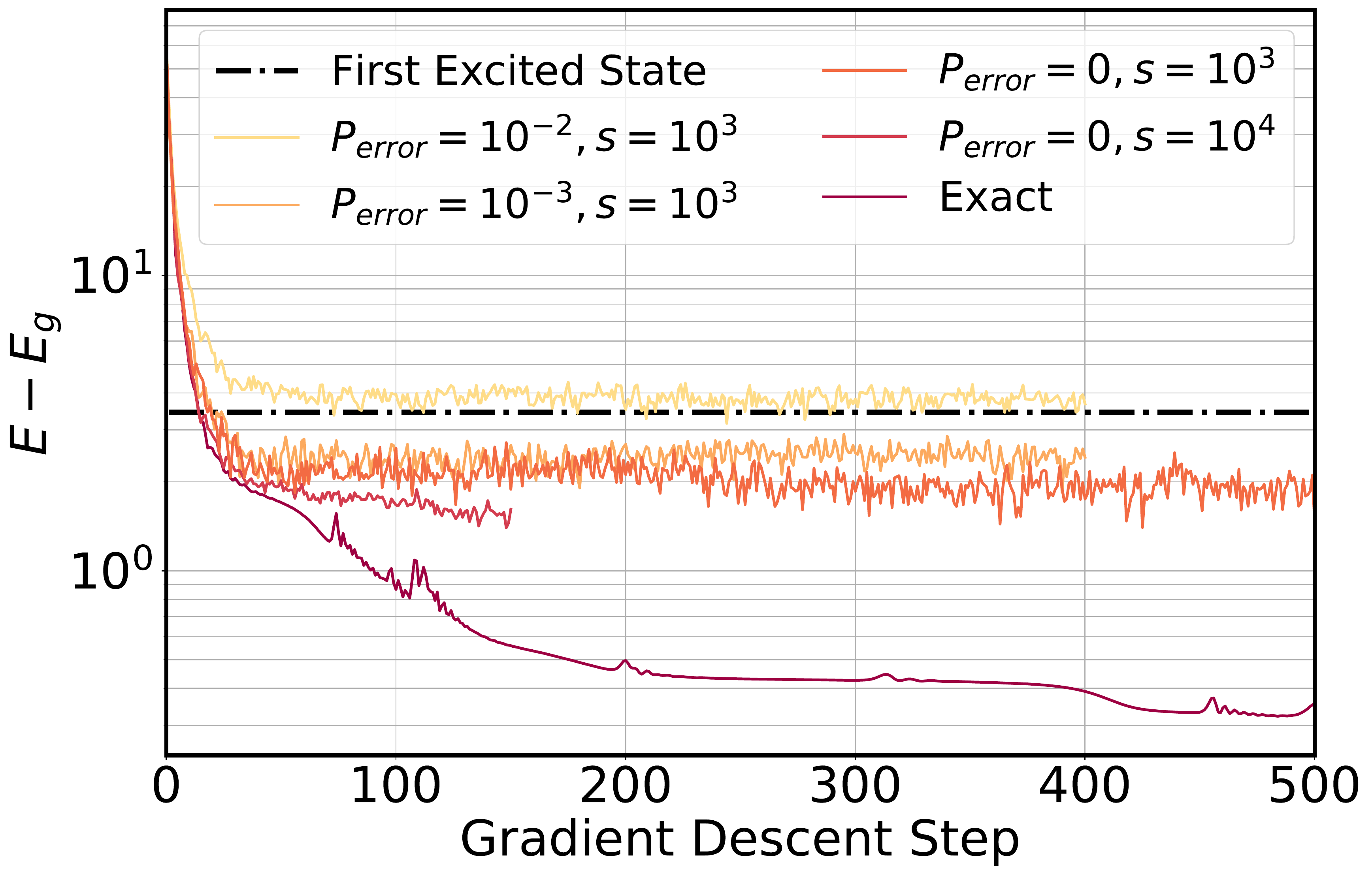}
\label{fig:noise_test-a}
}
\subfigure[][]{
\includegraphics[width=152pt]{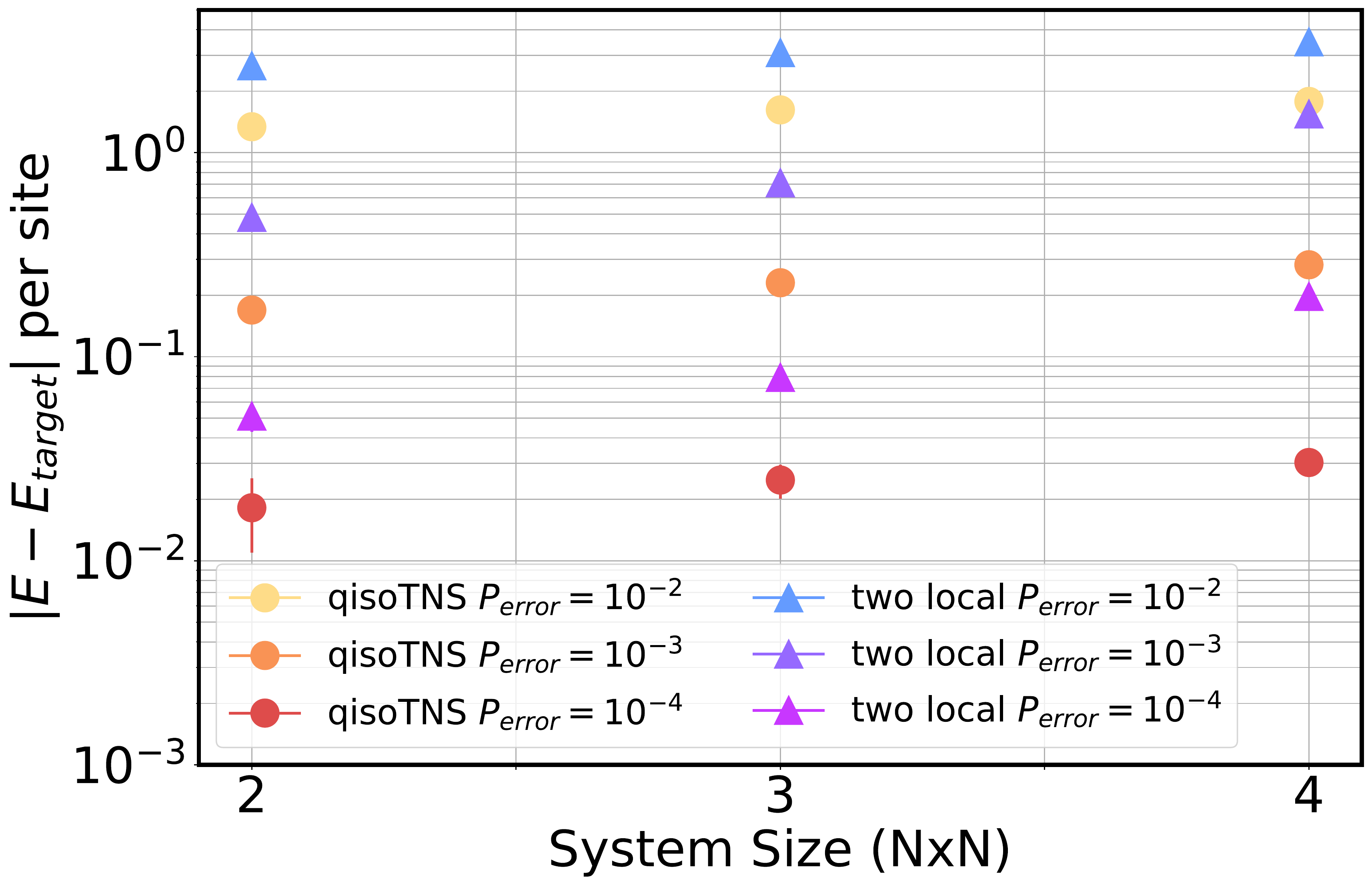}
\label{fig:noise_test-b}
}
\subfigure[][]{
\includegraphics[width=152pt]{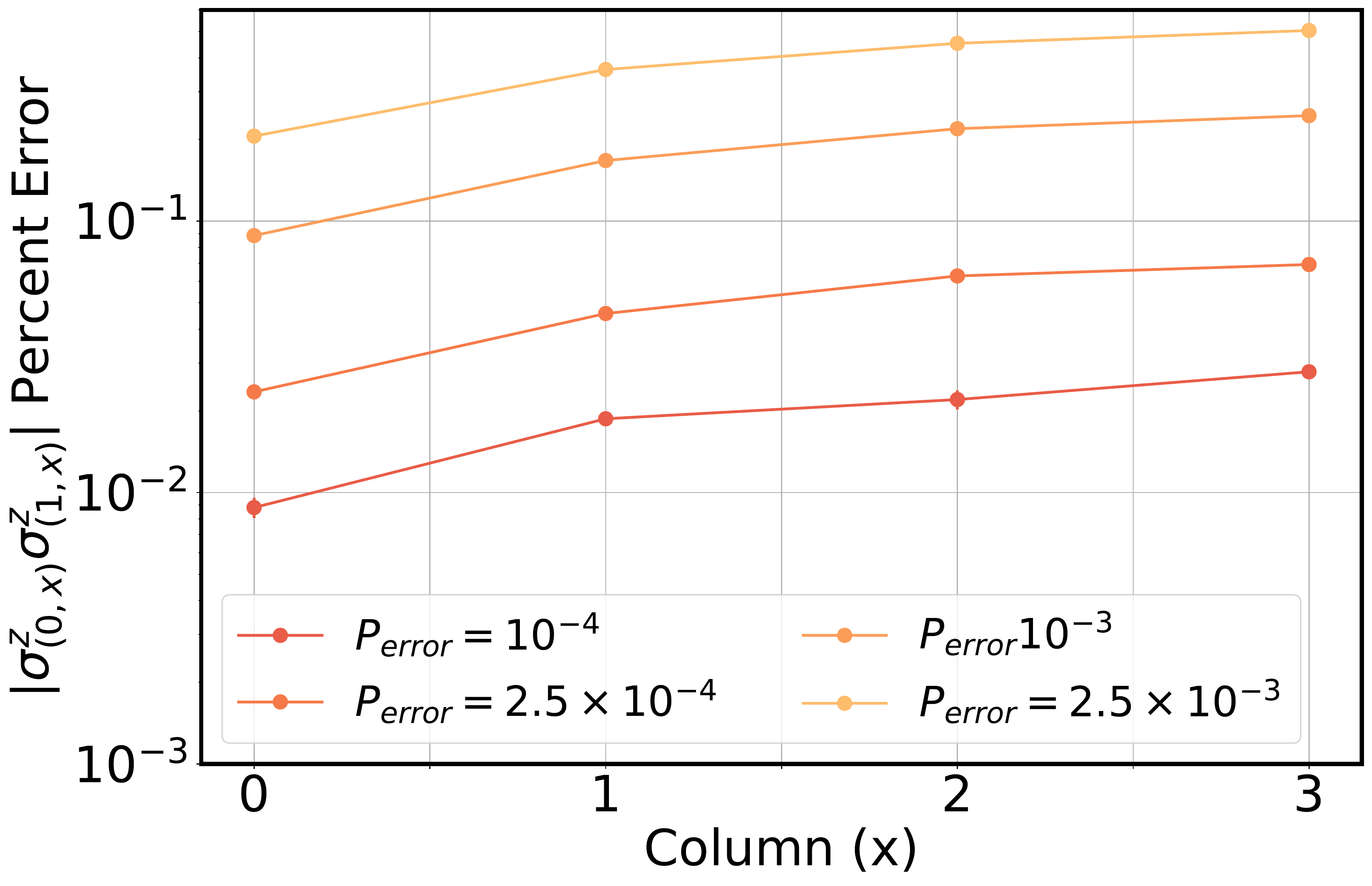}
\label{fig:noise_test-c}
}
\caption{\subref{fig:noise_test-a} VQE optimization of a $4 \times 4$  TFI model with $\Delta=1$ and $\lambda=3.5$ for a qisoTNS with $n_{bq}=1$ and $n_{bl}=3$ using both exact and noisy expectation value calculations. $P_{error}$ is the probability of either a depolarizing error or bit flip error occurring per gate. \subref{fig:noise_test-b} Per site energy difference from a target energy $E_{target}=E_g/(N^2)+0.1$  calculated using noisy quantum circuits. \subref{fig:noise_test-c} Percent error from exact values for $\sigma^{z}_{(0,x)}\sigma^{z}_{(1,x)}$ correlators calculated using noisy quantum circuits. Noisy expectation values are evaluated using Qiskit on a classical computer.}
\label{fig:noise_test}
\end{figure*}

In this section, we look at the performance of qisoTNS in the presence of noise.  We perform three tests. In the first test, we optimize a qisoTNS ansatz for a $4 \times 4$ TFI model (see Figure \ref{fig:noise_test-a}) under the presence of gate noise as well as stochastic expectation values coming from a finite (1000) number of samples.  We use one and two qubit gate depolarizing noise and bit flip errors. We find that stochastic and gate noise increase the final energy we plateau to (at least out to 500 gradient descent steps). 

In a second noise test, we look at the accuracy of computing the energy for two classes of ansatz: the qisoTNS and a standard hardware efficient ansatz made up of alternating layers of $U_3$ gates and CNOT entangling gates as in Figure \ref{fig:circ_comp-b}. For the same system size, each class of ansatz had approximately (within $10$ percent) the same number of $U_3$ and CNOT gates.  In both cases, we optimized (in a noise-free manner) to a target energy $E_{target}=E_{gs}+0.1N^2$ and then measured this energy under noisy conditions (both gate noise and stochastic noise). We used the same gate noise model as in the first test with 10,000 samples per expectation value.  We found that the qisoTNS consistently gave more accurate measurements for all $N$ and $P_{error}$ we studied (see Figure \ref{fig:noise_test-b}).  

Finally, we studied the measurement of $\sigma^z_{0,x}\sigma^z_{1,x}$ on the qisoTNS  under the stochastic and gate noise of Figure \ref{fig:noise_test-b}.  We find that the error from the exact result depends on the value of $x$ with correlators which involve sites which are deeper in the circuit being measured less accurately then shallow observables. This makes sense as observables which are deeper in the circuit have more time for errors from gates to accumulate.  This also helps us to further understand the second noise test as the standard hardware-efficient ansatz will uniformly pick up errors over all terms but the qisoTNS has some Hamiltonian terms that are significantly smaller errors (but still have some terms which will have large errors).

\section{\label{sec:conclusion}Conclusion}
In this paper, we introduced a quantum circuit version of the 2D isometric tensor network ansatz, qisoTNS. The qisoTNS ansatz is qubit efficient; the number of qubits necessary to simulate a $N \times N$ system scales as $O(N\log{\chi})$ instead of $O(N^2)$ as in a non-holographic variational ansatz. In addition, the qisoTNS ansatz is effectively trainable. Parameters nearest the orthogonality center remain trainable regardless of system size enabling one to employ a layerwise training strategy to ameliorate the barren plateau problem. We benchmarked the performance of the qisoTNS ansatz with two 2D spin 1/2 Hamiltonians. For the 2D TFI model on a $4 \times 4$ lattice, the ansatz with $n_{bq}=1$ was able to represent the ground state(s) well for all Hamiltonian parameters studied. In the case of the $J_1-J_2$ Heisenberg model on a $4 \times 4$ lattice with $J_1=1$ and $J_2=0.5$ near the model phase transition, we observed increased ground state fidelity for the $n_{bq}=2$ ansatz over the $n_{bq}=1$ ansatz indicating that for higher entanglement states, increasing the number of bond qubits can lead to better state fidelity.
In future work, it would be useful to further explore what models qisoTNS works well for. The 2D isometric tensor network has been used classically to simulate 2D spin 1/2 systems and string net liquids but applications to other systems are still unexplored \cite{Zaletel2020,Soejima2019,Haghshenas2019}; the exponential increase in accessible bond-dimension for qisoTNS will further expand simulatable states. In addition, modifications to qisoTNS in order to implement symmetries and different geometries would likely lead to more efficient training and representation and a fermionic version of the ansatz could extend the range of applicable models.

\begin{acknowledgements}
We acknowledge support from the Department of Energy grant DOE DE-SC0020165. This research is part of the Blue Waters sustained-petascale computing project, which is supported by the National Science Foundation (awards OCI-0725070 and ACI-1238993) the State of Illinois, and as of December, 2019, the National Geospatial-Intelligence Agency. Blue Waters is a joint effort of the University of Illinois at Urbana-Champaign and its National Center for Supercomputing Applications. This work made use of the Illinois Campus Cluster, a computing resource that is operated by the Illinois Campus Cluster Program (ICCP) in conjunction with the National Center for Supercomputing Applications (NCSA) and which is supported by funds from the University of Illinois at Urbana-Champaign.
\end{acknowledgements}

\bibliographystyle{apsrev4-1}
\bibliography{iso_paper}

\clearpage
\pagebreak
\input{appendix.tex}
\end{document}

%% file: appendix.tex
\appendix

\setcounter{section}{0}
\setcounter{equation}{0}
\counterwithout{equation}{section} 
\counterwithout*{equation}{section}
\renewcommand{\theequation}{A\arabic{equation}}
\renewcommand{\thesection}{\arabic{section}} 
\renewcommand\thefigure{A\arabic{figure}}  
\setcounter{figure}{0}   
\onecolumngrid

\section{\label{sec:AppA} Additional Isometric Tensor Networks}

\begin{figure}[h]
\subfigure[][]{
\includegraphics[height=210pt]{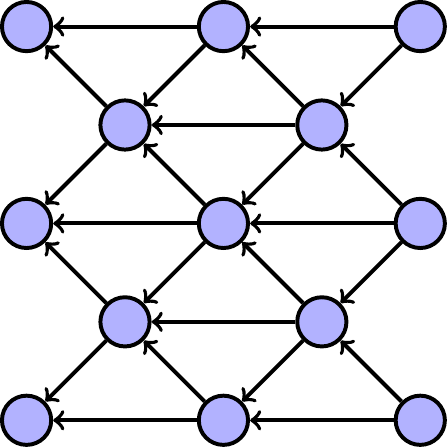}
\label{fig:tri-a}
}
\subfigure[][]{
\includegraphics[height=210pt]{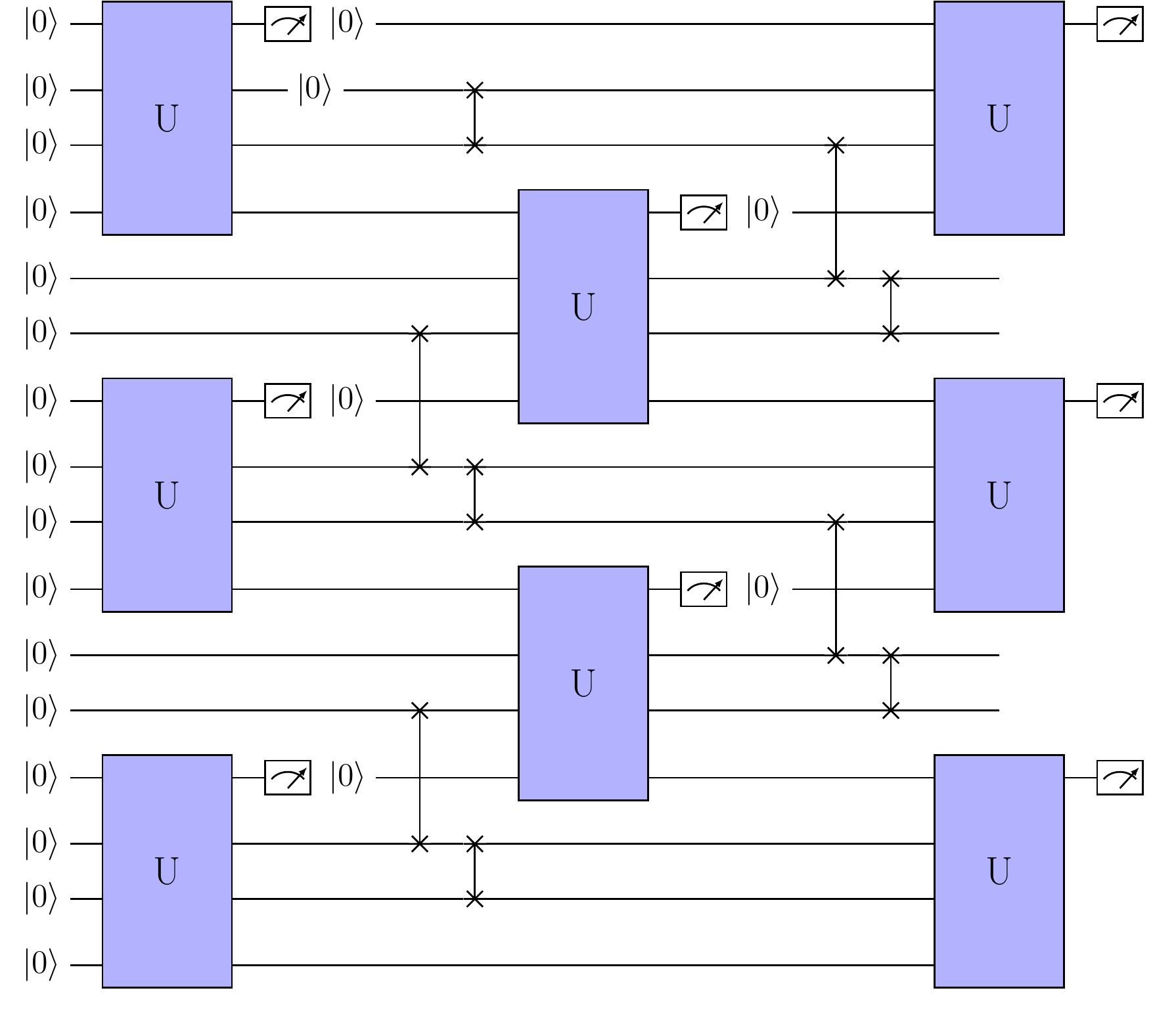}
\label{fig:tri-b}
}

\caption{\subref{fig:tri-a} A isometric tensor network for a 13 site triangular lattice. The arrows point in the direction of isometry. \subref{fig:tri-b} A $n_{bq}=1$ quantum circuit version of a 8 site triangular lattice.}
\label{fig:tri}
\end{figure}

\begin{figure}[h]
\subfigure[][]{
\includegraphics[height=160pt]{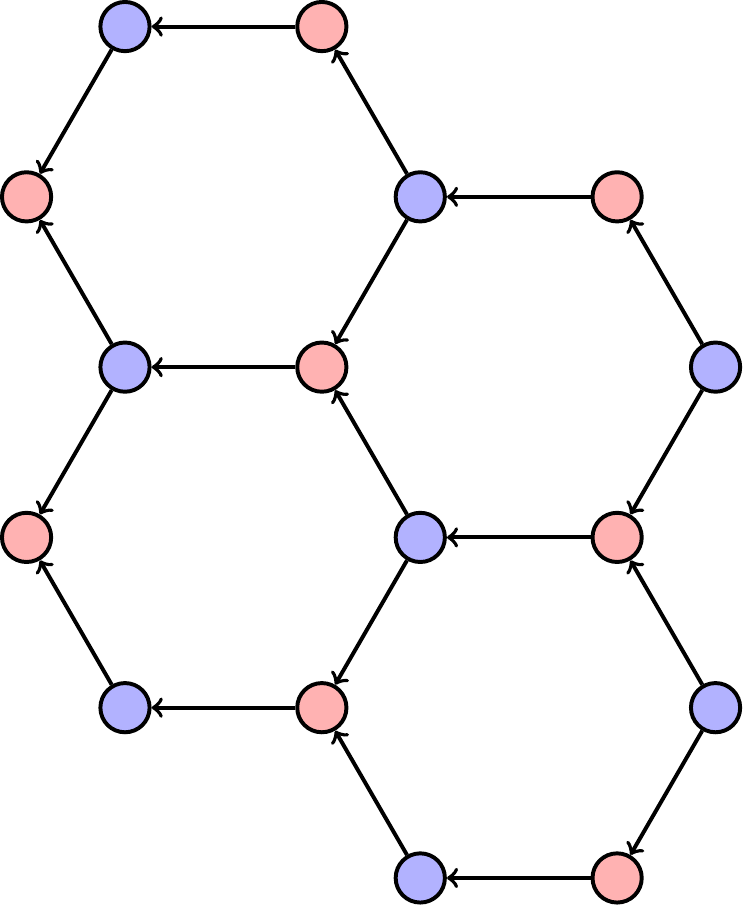}
\label{fig:honeycomb-a}
}
\subfigure[][]{
\includegraphics[height=160pt]{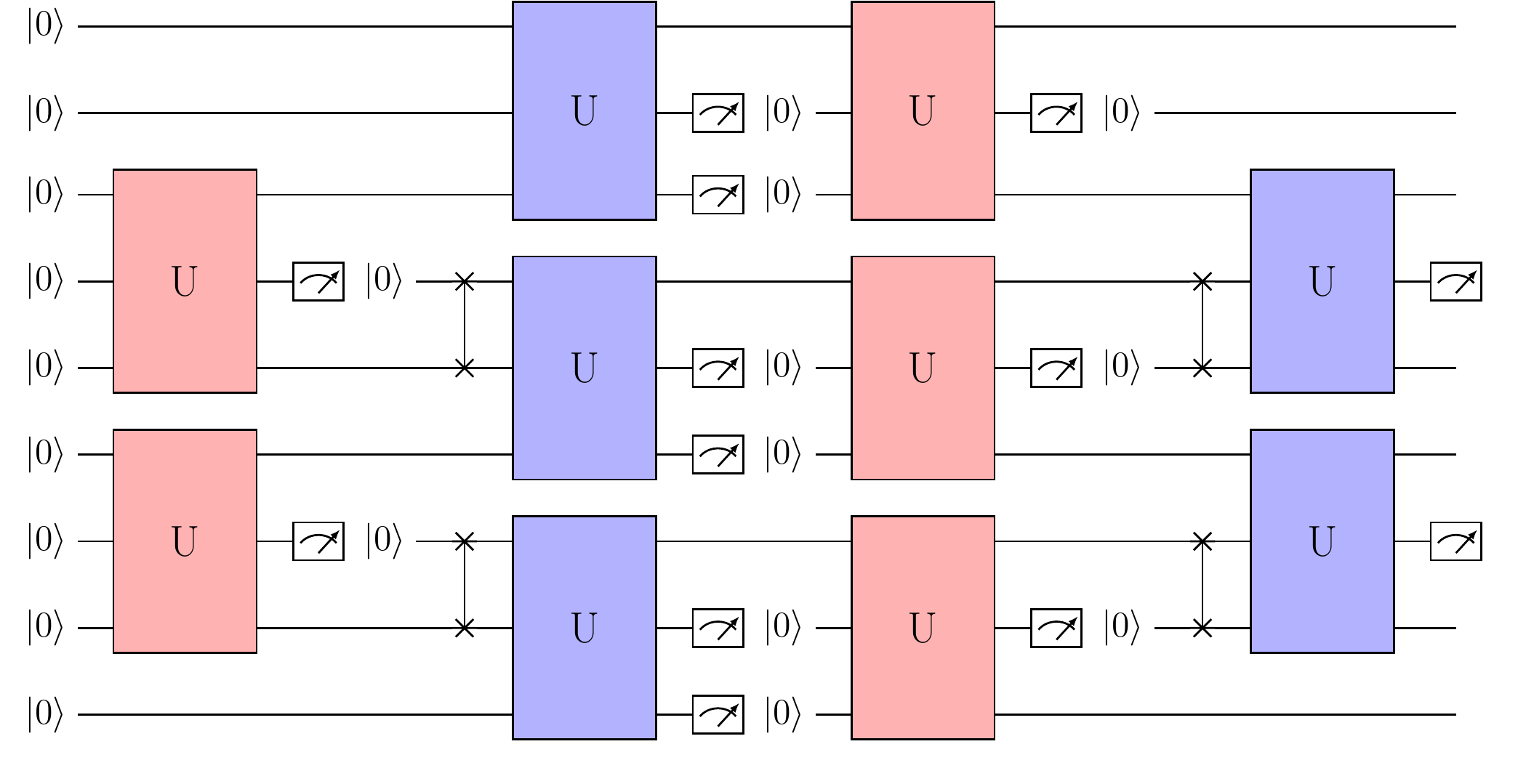}
\label{fig:honeycomb-b}
}

\caption{\subref{fig:tri-a} A isometric tensor network for a honeycomb lattice. The arrows point in the direction of isometry. In the bulk, the red tensors have two incoming and one outgoing arrow bonds while the blue tensors have one incoming and two outgoing arrow bonds. \subref{fig:tri-b} A $n_{bq}=1$ quantum circuit version of a 10 site honeycomb lattice. For this circuit, the middle site of a tensor represents the physical index. For the blue tensors, because the have only one incoming arrows bonds, we reset the bottom qubit in order to not carry entanglement forward in the circuit.}
\label{fig:honeycomb}
\end{figure}

While in the main body of this work we examined the isometric generalization of PEPs, other 2D tensor network structures are possible. In Figure \ref{fig:tri-a}, we show how to construct an isometric tensor network for a triangular lattice. Each lattice site in the bulk, has 3 incoming and 3 outgoing bonds. In Figure \ref{fig:honeycomb-a}, we show how to construct an isometric tensor network for a honeycomb lattice. Each lattice site, in the bulk, has either  two incoming and one outgoing bonds or one incoming and two outgoing bonds.  Note that for both of these lattices there are no orthogonality centers or hypersurfaces making classical computation of expectation values intractable. In Figures \ref{fig:tri-b} and \ref{fig:honeycomb-b}, we show the quantum circuit version of the triangular lattice and honeycomb lattice respectively. As in the square lattice case, we use swap gates to connect outgoing wires with the appropriate incoming wires of the next gate.

